\pdfoutput=1
\documentclass[11pt,twoside,a4paper,cmspaper,final,collab]{cms-tdr}
\def\svnVersion{5cb61d7-D}\def\svnDate{2020/11/03}\def\cmsCernNoTag{CERN-EP-2020-113}\def\cmsCernDate{\today}\def\cmsMessage{"Published in the Journal of High Energy Physics as \href{http://dx.doi.org/10.1007/JHEP01(2021)163}{\doi{10.1007/JHEP01(2021)163}.}"}
\begin{document}\cmsNoteHeader{BPH-17-004}

\newcommand{\tautrimu}{\ensuremath{\PGt \!\to\! 3\PGm}\xspace}
\newcommand{\Wtaunu}{\ensuremath{\PW \!\to\! \PGt\PGn}\xspace}
\newcommand{\Wmunu}{\ensuremath{\PW \!\to\! \PGm\PGn}\xspace}
\newcommand{\ppWX}{\ensuremath{\Pp\Pp \!\to\! \PW + X}\xspace}
\newcommand{\dstaunu}{\ensuremath{\PsDp \!\to\! \PGtp\PGn}\xspace}
\newcommand{\dsphipi}{\ensuremath{\PsDp \!\to\! \PGf\PGpp \!\to\! \PGmp\PGmm\PGpp}\xspace}
\newcommand{\BtauX}{\ensuremath{\PB \!\to\! \PGt + X}\xspace}
\newcommand{\BDsX}{\ensuremath{\PB \!\to\! \PsDp + X}\xspace}
\newcommand{\ptvecmumumu}{\ensuremath{\ptvec^{\kern0.7pt{3\PGm}}}\xspace}
\newcommand{\ptmumumu}{\ensuremath{\pt^{\kern0.7pt{3\PGm}}}\xspace}
\DeclareRobustCommand{\PBsz}{\HepParticle{\PB}{s}{0}\xspace}

\cmsNoteHeader{BPH-17-004}
\title{Search for the lepton flavor violating decay \texorpdfstring{\tautrimu}{tau to 3mu} in proton-proton collisions at $\sqrt{s} = 13\TeV$}

\author{The CMS Collaboration}

\date{\today}

\abstract{
Results are reported from a search for the lepton flavor violating decay \tautrimu in proton-proton
collisions at $\sqrt{s}=13\TeV$. The data sample corresponds to an integrated luminosity of 33.2\fbinv recorded by the
CMS experiment at the LHC in 2016. The search exploits \PGt leptons produced in both \PW boson and heavy-flavor hadron
decays. No significant excess above the expected background is observed. An upper limit on
the branching fraction $\mathcal{B}(\tautrimu)$ of $8.0 \times 10^{-8}$ at 90\% confidence level is obtained, with an expected upper
limit of $6.9 \times 10^{-8}$.
}

\hypersetup{%
pdfauthor={CMS Collaboration},%
pdftitle={Search for the lepton flavor violating decay tau to 3mu in proton-proton collisions at sqrt(s) = 13 TeV},%
pdfsubject={CMS},%
pdfkeywords={CMS, physics, BSM, CLFV, charged lepton flavor violation}}

\maketitle

\section{Introduction}
\label{sec:intro}
In the standard model (SM) with massless neutrinos, the three lepton flavor numbers are exactly conserved.  The
observation of neutrino oscillations not only proves that lepton flavor is not conserved in the neutral sector, but also
provides a mechanism, through neutrino loops, for lepton flavor violating (LFV) decays of charged leptons such as \tautrimu,
albeit with extraordinarily small
branching fractions~\cite{Pham:1998fq,Hernandez-Tome:2018fbq,Blackstone:2019njl}.  However, a number of SM extensions predict a
much larger \tautrimu branching
fraction, including values as high as $10^{-10}$--$10^{-8}$~\cite{Marciano:2008zz,Raidal:2008jk,Arganda:2005ji}, accessible
to current and near-future experiments.  The BaBar Collaboration set a limit of $\mathcal{B}(\tautrimu) < 5.3 \times 10^{-8}$ at
90\% confidence level (\CL)~\cite{Aubert:2007pw}.  The present best limit of ${<}2.1 \times 10^{-8}$ at 90\% CL was obtained by the
Belle experiment~\cite{Hayasaka:2010np}.  Searches at the CERN LHC
are approaching this sensitivity with 90\% \CL upper limits of $4.6\times 10^{-8}$ from
LHCb~\cite{Aaij:2014azz} and $38\times 10^{-8}$ from ATLAS~\cite{Aad:2016wce}.

The LHCb and ATLAS results targeted \PGt production from heavy-flavor hadron decays and \PW boson decays,
respectively.  While many more \PGt leptons are produced from heavy-flavor hadron decays, the \PGt leptons from \PW decays tend
to have larger transverse momentum (\pt) and are typically isolated from hadronic activity, providing an experimental signature
with much less background.  In this paper, we present results from the CMS experiment of the first search for the LFV decay \tautrimu from a
combination of the two independent channels (production in \PW boson and heavy-flavor hadron decays).  Using both channels, for which
CMS has comparable sensitivity, provides the best opportunity for a discovery or the lowest upper limit on the branching fraction.
The data were collected at the LHC in 2016 from proton-proton ($\Pp\Pp$) collisions at a center-of-mass
energy of 13\TeV, and correspond to an integrated luminosity of 33.2\fbinv. Inclusion of charge-conjugate states is implied
throughout this paper.

\section{The CMS experiment}
\label{sec:CMS}
The central feature of the CMS apparatus is a superconducting solenoid of 6\unit{m} internal diameter, providing a
magnetic field of 3.8\unit{T}. Within the solenoid volume are a silicon pixel and strip tracker, a lead
tungstate crystal electromagnetic calorimeter, and a brass and scintillator hadron calorimeter, each composed of a
barrel and two endcap sections. Muons are measured in gas-ionization detectors embedded in the steel flux-return yoke
outside the solenoid. Additional forward calorimetry complements the coverage provided by the barrel and endcap
detectors. A more detailed description of the CMS detector, together with a definition of the coordinate system used and
the relevant kinematic variables, can be found in Ref.~\cite{Chatrchyan:2008zzk}.

Events of interest are selected using a two-tiered trigger system~\cite{Khachatryan:2016bia}. The first level (L1),
composed of custom hardware processors, uses information from the calorimeters and muon detectors to select events at a
rate of around 100\unit{kHz} within a fixed time interval of less than 4\mus. The second level, known as the high-level
trigger, consists of a farm of processors running a version of the full event reconstruction software optimized
for fast processing, and reduces the event rate to around 1\unit{kHz} before data storage.

The particle-flow algorithm~\cite{Sirunyan:2017ulk} aims to reconstruct and identify each individual particle in an
event, with an optimized combination of information from the various elements of the CMS detector. In particular, muons
are identified by matching tracks in the silicon tracker with tracks in the muon detector and verifying the energy
deposited in the calorimeters is consistent with that expected for muons.  The muon momentum is obtained from the
curvature observed in the silicon tracker and the relative \pt resolution for muons with $\pt<100\GeV$ is 1\% in the
barrel and 3\% in the endcaps~\cite{Sirunyan:2018fpa}.

Simulated event samples are used to validate the analysis, measure acceptance and efficiency, and estimate systematic
uncertainties.
For the analysis of \PGt leptons from \PW boson decays, events were simulated using \MGvATNLO 2.5.2~\cite{Frixione:2002ik, Alwall:2014hca}
at leading order, assuming a two-Higgs-doublet model that allows for flavor changing neutral currents and LFV processes, interfaced
with \PYTHIA for parton shower and hadronization descriptions. The \PW production and decay, as well as the \PGt decay, are handled by \MGvATNLO. The \pt
distribution of the \PW boson is reweighted to match that obtained from a SM next-to-leading-order $\PW \!\to\! \ell \PGn$ sample
produced with \MGvATNLO and interfaced to \PYTHIA for parton showers and hadronization. 
For the analysis of \PGt leptons from heavy-flavor hadron decays, events were simulated
using \PYTHIA8.226~\cite{PYTHIA} with the CUETP8M1 tune~\cite{Khachatryan:2015pea} interfaced with \EVTGEN1.6.0~\cite{EVTGEN} for particle
decays, with the \PGt decay kinematics determined by phase space, rather than a particular model.
All events are passed through the CMS detector simulation based
on \GEANTfour~\cite{GEANT}.  The multiple $\Pp\Pp$ collisions that occur within the same or nearby bunch crossings (pileup) are modeled by
including additional minimum bias events generated with \PYTHIA with a distribution that matches the one observed in data.  Simulated
events are reconstructed with the same algorithms as used for data, including emulation of the triggers.

\section{Data selection}
\label{sec:selection}
The triggers used by this analysis evolved during the
data collection period, primarily to cope with increases in the instantaneous luminosity.  Most of the data were
collected with an L1 trigger requirement of either three muons, two muons with at least one muon
having $\pt>10\GeV$, or two muons with both muons having an absolute pseudorapidity of $\abs{\eta}<1.6$.  The dimuon L1
triggers also required the two muons to have an absolute pseudorapidity difference $\abs{\Delta\eta}<1.8$.  The high-level trigger required three reconstructed
charged particles (tracks), of which two must be identified as muons with $\pt>3\GeV$ and the other must have
$\pt>1.2\GeV$.  The three tracks are fitted to a common vertex and kept if the normalized $\chi^2$ of the fit is less than
8; the vertex location is at least 2 times its uncertainty from the beamline; the \pt of the combination (\ptmumumu) is
greater than 8\GeV; the invariant mass (assuming a muon mass for all tracks) is in the range 1.60--2.02\GeV; and the cosine of
the angle in the transverse plane between the three-track momentum vector and the vector from the beamline to the vertex is greater
than 0.9.  During the first half of 2016, errors in the L1 triggers used by this analysis resulted in a significant
loss of efficiency for muons with $\abs{\eta}>1.24$.  While this trigger misconfiguration is not modeled by the simulation,
it is accounted for by the analysis.

Offline, all combinations of three muons in the event with a combined charge of ${\pm}1$ are considered and a fit to a
common vertex is attempted to make a \PGt candidate.  The muons are required to match the ones used in the trigger, and
the trigger-level selection criteria are reapplied.  If either of the oppositely charged dimuon combinations from
the \PGt candidate has an invariant mass within 20\MeV of the mass of the \Pgo or \Pgf resonances, the candidate is
rejected.  Events with at least one \PGt candidate with an invariant mass between 1.6 and 2.0\GeV are kept for analysis
by two different algorithms, one optimized for production of \PGt leptons in \PW boson decays and the other optimized
for production of \PGt leptons in heavy-flavor hadron decays.

\section{Search for \texorpdfstring{\tautrimu}{tau to 3mu} in \texorpdfstring{\PW}{W} boson decays}
\label{sec:Wsearch}

\subsection{Selecting \texorpdfstring{$\PGt$}{tau} candidates}
\label{ssec:Wselection}
For the \PW boson analysis, \PGt candidates must pass the selection criteria described in Section~\ref{sec:selection}, as well as
an additional veto that suppresses background arising from dimuon decays of hadronic resonances.
The veto considers all pairs of oppositely charged muons with one muon from the \PGt candidate and
one muon not associated with the \PGt candidate.  If any of the muon pairs form a good vertex (vertex fit $\chi^2$
probability above 5\%) and have an invariant mass within twice the larger of the detector resolution or natural width of a known
resonance with a dimuon decay, the \PGt candidate is vetoed.  The checked resonances
are \Pgh, \Pgo, \Pgr, \Pgf, \PJGy, \Pgy, \PgUa, \PgUb, \PgUc, and \PZ.

The reconstructed $\Pp\Pp$ interaction vertex with the largest value of summed physics-object $\pt^2$ is taken to be the primary interaction  vertex.
The physics objects are the jets, clustered using the anti-\kt
algorithm~\cite{Cacciari:2008gp,Cacciari:2011ma} with the tracks assigned to candidate vertices as inputs, and the
associated missing transverse momentum, taken as the negative vector sum of the \ptvec of those jets.

To better separate signal from background, a boosted decision tree (BDT) is trained~\cite{Chen:2016:XST:2939672.2939785}
using simulated signal events and background from data events in the mass sideband region (trimuon invariant mass
in the range of 1.60--1.74 or 1.82--2.00\GeV).  The signal sample used for training the BDT is a combination of several samples, each
with a different \PGt lepton mass (covering the mass range 1.6--2\GeV), to avoid training on the true \PGt mass.
Data in the mass sidebands contain combinatorial
background, as well as decays, primarily of heavy-flavor hadrons, where one or more hadrons are misreconstructed as muons.  Simulated data were used to
verify that background from charm hadron decays does not produce a peak in the trimuon mass.

The BDT uses 18 variables.  The variables include a measure of the muon quality for each muon, the difference
in longitudinal impact parameters with respect to the primary vertex for each pair of muons,
the \pt and $\eta$ of the \PGt candidate, the $\chi^2$ of the
trimuon vertex fit, the distance in the transverse plane between the trimuon vertex and the
beamline divided by the uncertainty in that distance, and the angle in the transverse plane between the trimuon momentum vector and the vector between the beamline and the
trimuon vertex.  The remaining variables include additional information about the event.  The absolute isolation of
the \PGt candidate~\cite{Sirunyan:2018pgf} is the sum of the transverse momenta of the charged particles (charged isolation) and photons
(neutral isolation) reconstructed using the particle-flow algorithm, with $\Delta R = \sqrt{\smash[b]{(\Delta\eta)^2 + (\Delta\phi)^2}} < 0.5$,
where $\Delta\eta$ and $\Delta\phi$ are the differences in pseudorapidity and azimuthal angle, respectively,
between the directions of the particle and the \PGt candidate.  The charged isolation only includes tracks that pass
within 0.2\cm of the primary vertex in the longitudinal direction and are not one of the \PGt candidate constituents.
The neutral isolation is corrected for pileup following the prescription in Ref.~\cite{Sirunyan:2018pgf}.
The variable used in the BDT is the relative isolation, defined as the absolute isolation divided by the \pt of the \PGt candidate.

Assuming that the only missing particle in the event
is the neutrino from the \Wtaunu decay, the neutrino \ptvec can be determined from the negative vector sum
of the transverse momenta of all other particles in the event, a quantity referred to as \ptvecmiss. A multivariate
regression that uses additional information from the event~\cite{CMS-DP-2015-042} is applied to \ptvecmiss to reduce
effects from pileup, improving the \ptvecmiss resolution by 30\%.  The \PW boson \ptvec is defined as the sum of \ptvecmiss
and \ptvecmumumu.  Furthermore, using the known mass of the \PW boson, the longitudinal momentum of the neutrino can be
determined, up to a two-fold ambiguity.  The remaining BDT variables use this information and are: both neutrino
longitudinal momentum solutions, \PW boson \pt, \ptmiss, the angle $\Delta\phi$ in the transverse plane
between \ptvecmiss and \ptvecmumumu, and the transverse \PW mass $\sqrt{\smash[b]{2\ptmumumu\ptmiss (1 - \cos{\Delta\phi})}}$.

The BDT is trained and tested on independent samples with no evidence of overtraining or bias. The
most important variables are found to be the \PGt candidate relative isolation, transverse \PW mass, and \ptmumumu.

\subsection{Analysis strategy}
\label{ssec:Wanalysis}

The relationship between the \tautrimu branching fraction and the number of signal events can be written as:
\begin{linenomath}
\begin{equation}
\label{eq:wbr}
\mathcal{B}(\tautrimu) = \frac{N_{\text{sig}(\PW)}}
{\mathcal{L}\, \sigma(\ppWX) \, \mathcal{B}(\Wtaunu)\,
\mathcal{A}_{3\Pgm(\PW)}\, \epsilon_{3\PGm(\PW)}},
\end{equation}
\end{linenomath}
where $N_{\text{sig}(\PW)}$ is the number of signal events, $\mathcal{L}$ is the integrated luminosity,
$\sigma(\ppWX)$ is the \PW boson production cross section, $\mathcal{B}(\Wtaunu)$ is the branching
fraction of \PW decay to $\PGt\PGn$, $\mathcal{A}_{3\Pgm(\PW)}$ is the acceptance, and $\epsilon_{3\PGm(\PW)}$ is
the combined reconstruction, selection, and trigger efficiency for the three muons.  The product of $\sigma(\ppWX)$ and
$\mathcal{B}(\Wtaunu)$ is obtained from the ATLAS measurement of
$\sigma(\ppWX) \mathcal{B}(\Wmunu)$ at 13 TeV~\cite{Aad:2016naf} and the world-average value of the ratio
$\mathcal{B}(\Wtaunu) / \mathcal{B}(\Wmunu)$~\cite{PDG2018}.
Other sources of \PGt leptons, such as from \PZ boson or \PD meson decays, are neglected as either the low production 
cross section or BDT selection efficiency reduces their contribution to no more than a few percent of that from \PW boson production.

Simulated samples are used to estimate the relative production of \PGt leptons from different sources and to determine
the acceptance and efficiency of the signal.  To account for differences between data and simulation, several
multiplicative corrections are applied on an event-by-event basis to the simulated events.  Each of the three muons has
a weight associated with it, which is the product of three corrections related to the efficiency of reconstructing the
track in the tracker, the efficiency of identifying the reconstructed track as a muon, and the efficiency for the
trigger system to find the muon given that it was reconstructed and identified by the offline algorithm.  An additional
correction is applied to account for the L1 trigger misconfiguration described in Section~\ref{sec:selection}.
The average weight from the combination of these corrections is 0.88.
The difference from unity comes primarily from the
trigger efficiency.  The weighted events are used to determine the signal efficiency, and the uncertainties
from the corrections are included as systematic uncertainties.

\begin{figure}[hbtp]
  \centering
    \includegraphics[width=0.45\textwidth]{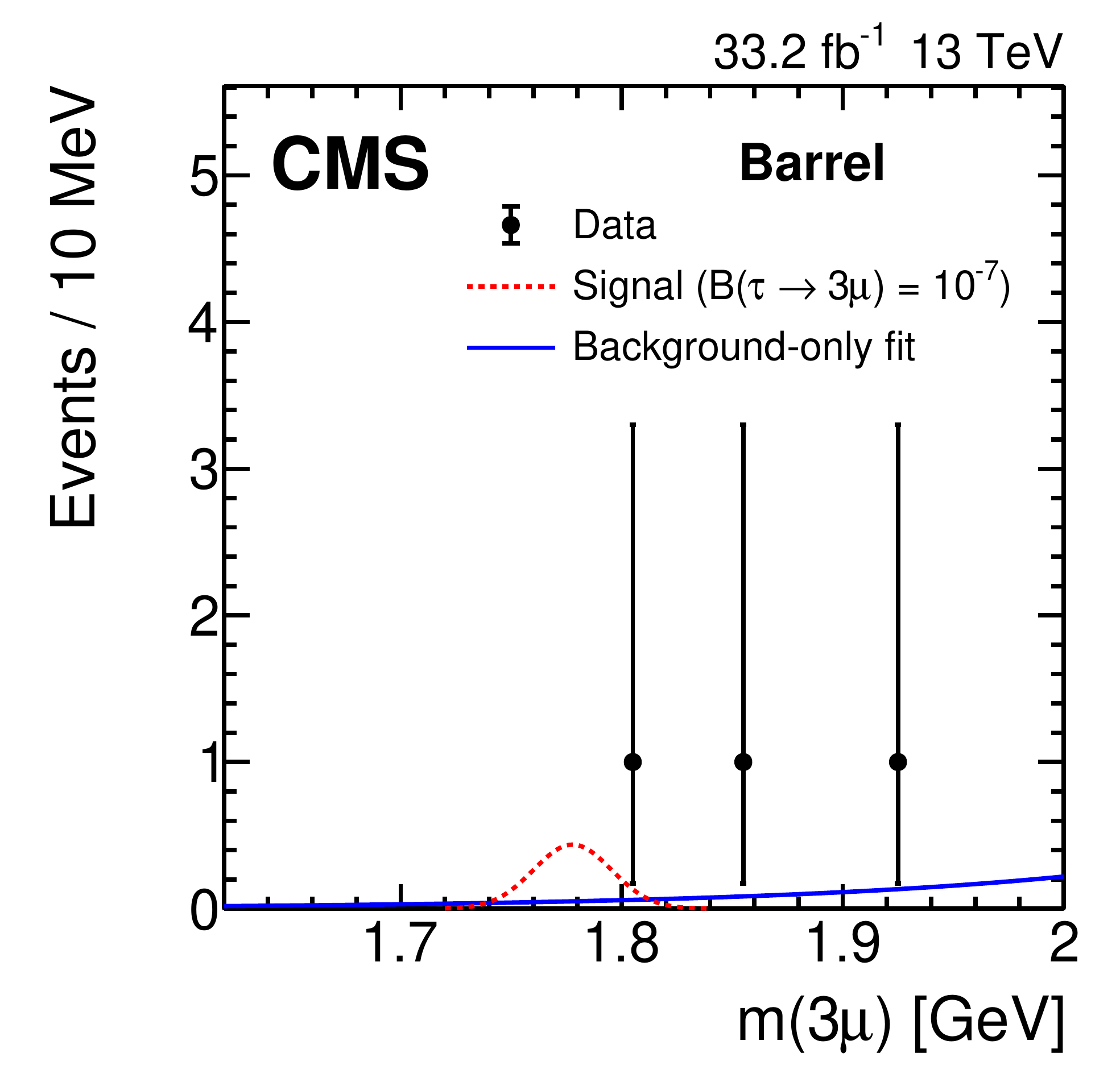}
    \includegraphics[width=0.45\textwidth]{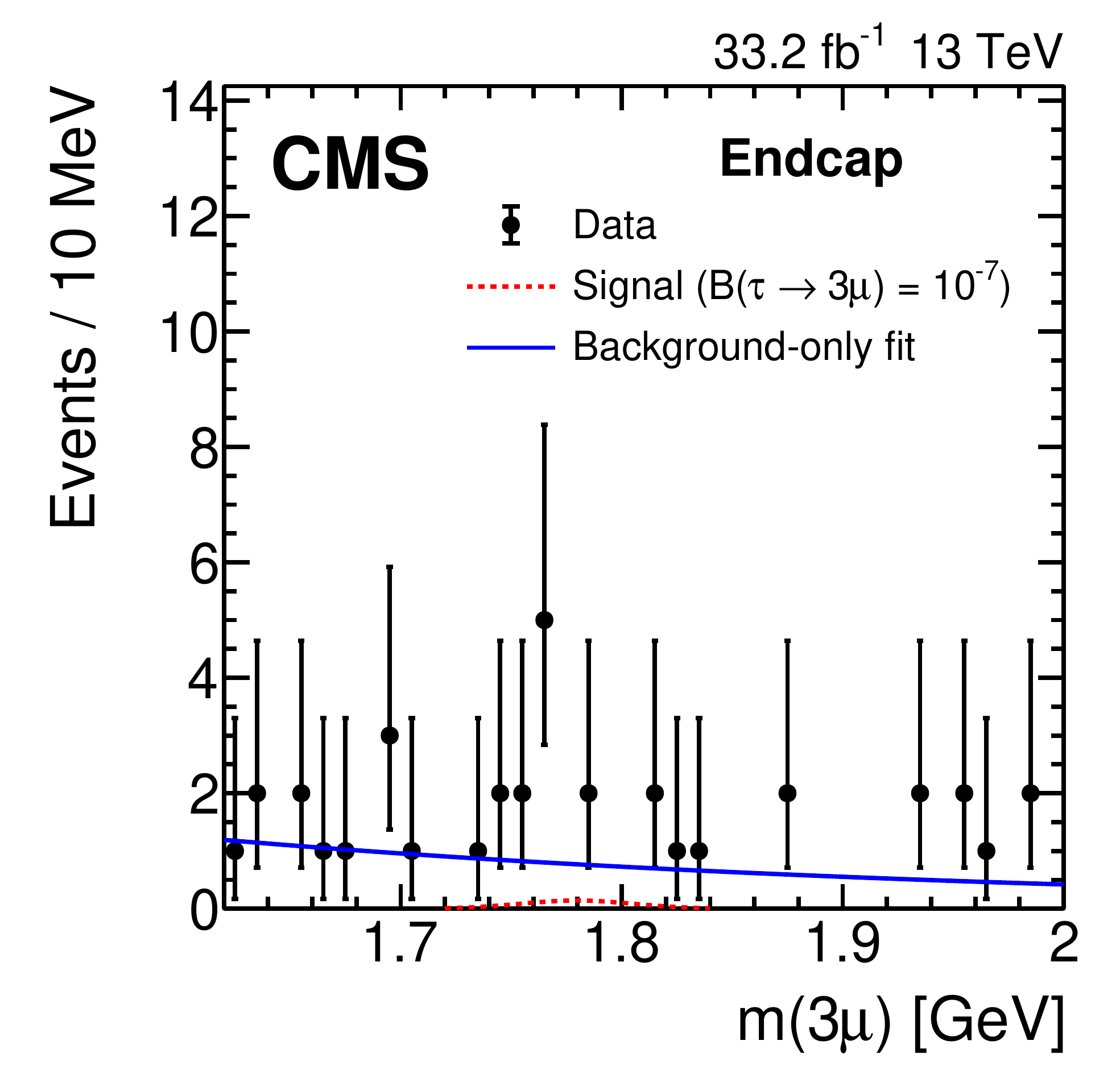}
 \caption{Trimuon invariant mass distributions for barrel (left) and endcap (right) categories of the \PW boson analysis.
 The data are shown with filled circles and vertical bars representing the statistical
uncertainty.  The background-only fit and the expected signal for $\mathcal{B}(\tautrimu) = 10^{-7}$ are shown with
solid and dashed lines, respectively.}
\label{fig:W_massplots}
\end{figure}

Since the $\PGt$ invariant mass resolution is a strong function of the $\PGt$ pseudorapidity, the data sample is
divided into two mutually exclusive categories, barrel and endcap, corresponding to trimuon $\abs{\eta}<1.6$ (with an average
mass resolution of 16\MeV) and $\abs{\eta}\geq1.6$ (with an average mass resolution of 27\MeV), respectively.  Events with a BDT score larger than a
given threshold are selected and used for the final analysis. Simulated signal and sideband data events are used to
set the BDT score thresholds for the barrel and endcap regions that give the most stringent expected exclusion limits.
Figure~\ref{fig:W_massplots} shows the trimuon invariant mass distributions for events passing each category, along
with a background-only fit (described in Section~\ref{sec:results}) and the contribution expected for a signal with $\mathcal{B}(\tautrimu) = 10^{-7}$.

\subsection{Systematic uncertainties}
\label{ssec:Wsystematics}
The largest systematic uncertainty is from the corrections that are used in extracting the
signal efficiency.  This is dominated by the L1 trigger inefficiency correction, which predominantly affects the endcap
region, and is correlated between the barrel and endcap categories.  The other simulation correction uncertainties are
uncorrelated between the two categories.  The second largest systematic uncertainty arises from the limited size of the
simulated samples and is uncorrelated between the two categories.  The remaining uncertainties come from the integrated
luminosity~\cite{CMS-PAS-LUM-17-001}, the \PW boson production cross section, and the \PW boson branching fractions, all
of which are correlated between the barrel and endcap categories.  The systematic uncertainties are summarized in
Table~\ref{tab:Wsystematics}.

\begin{table}[htbp]
\centering
\topcaption{Sources of systematic uncertainties in the \PW boson analysis and their effect on the signal efficiency and
normalization for the barrel and endcap categories.}
\begin{tabular}{lcc}
 & \multicolumn{2}{c}{Uncertainty (\%)} \\
Source & Barrel & Endcap \\
\hline
Signal efficiency                            &  7.9       & 32      \\ 
Limited size of simulated samples &  4.3       &  6.2      \\
Integrated luminosity                        &  2.5       &  2.5      \\
$\Pp\Pp\!\to\! \PW$ cross section                    &  2.9       &  2.9      \\
$\mathcal{B} (\Wmunu)$             &  0.2       &  0.2      \\
$\mathcal{B} (\Wtaunu)$             &  0.2       &  0.2      \\
\hline
Total                            &  9.8       & 33      \\
\end{tabular}
\label{tab:Wsystematics}
\end{table}

\section{Search for \texorpdfstring{\tautrimu}{tau to 3mu} in heavy-flavor hadron decays}
\label{sec:HFsearch}
The measurement of the \tautrimu branching fraction for \PGt leptons produced in charm and bottom decays is complicated
by uncertainties in the production of heavy-flavor hadrons.  These uncertainties are reduced by utilizing the
decay \dsphipi to normalize the signal yield.

Simulated samples are used to estimate the relative production of \PGt leptons from different sources and to determine
the acceptance and efficiency of the signal and normalization modes.  Four samples are used to extract the acceptance and
efficiency.  The first is a sample of \dstaunu decays.  The second and third samples contain the inclusive
$\PBp\!\to\!\PGt +X$ and $\PBz\!\to\!\PGt+ X$ decays, respectively.  The fourth sample contains \dsphipi events.  For all
samples, the heavy-flavor decays are simulated with \EVTGEN1.6.0, with the \tautrimu decay occurring via phase space.  In
the first and fourth samples, the \PsDp mesons can be produced by hadronization or from \PQb hadron decays.  The
acceptance $\mathcal{A}$ is the fraction of events in which all tracks of the \PGt or \PsDp decay have $\abs{\eta}<2.4$,
the muons have $p>2.5\GeV$, and the pion (if present) has $\pt>1\GeV$.  The efficiency is the product of the
reconstruction and selection efficiency $\epsilon^\text{reco}$ and the trigger efficiency
$\epsilon^\text{trig}$.

\subsection{Selecting \texorpdfstring{$\tau$}{tau} candidates}
\label{ssec:HFselection}
For the heavy-flavor analysis, \PGt candidates must pass the selection criteria described in Section~\ref{sec:selection}
and the lowest-\pt track must have $\pt>2\GeV$.  The trimuon sample is divided into a signal region (invariant mass of
1.75--1.80\GeV) and a sideband (background) region (invariant mass of 1.60--1.75 or 1.80--2.00\GeV).  The normalization
channel \dsphipi uses the same selection criteria with a few exceptions.  Only two
muons are required and they must be oppositely charged with an invariant mass between 1 and 1.04\GeV.  The track
associated with the pion must have $\pt>2\GeV$ and form a vertex with the two muons with a normalized $\chi^2$ less than
5.  The three-track invariant mass must be in the range 1.68--2.02\GeV, with the signal region defined as
1.93--2.01\GeV and the sideband region as 1.70--1.80\GeV.  If there is more than one \PGt or $\PsDp$ candidate in an
event, the one with the smallest vertex fit $\chi^2$ is selected.  Once a candidate is found, its trajectory is
extrapolated to the beamline and the primary vertex is selected as the reconstructed $\Pp\Pp$ collision vertex that is closest
to the extrapolated point.

To improve the signal-to-background ratio for the \tautrimu sample, a BDT is trained using simulated signal events
(including $\PGt$ leptons produced from both charm and bottom decays)
and background events from the data sideband region.
The training utilizes 10 variables: the smallest muon momentum, three distinct muon quality criteria (each using the
``worst'' value of the three muon candidates), the $\chi^2$ of the trimuon vertex fit, the angle between the trimuon momentum
vector and the vector connecting the primary and trimuon vertices, the distance between the
trimuon vertex and the primary vertex divided by the uncertainty in that distance, the smallest transverse impact parameter of the muons with respect to the
primary vertex, and two isolation variables.  The first isolation variable is the smallest distance of closest approach to the trimuon vertex of all
other tracks in the event with $\pt>1\GeV$.  The second isolation variable sums the \pt of all tracks with
$\pt>1\GeV$, $\Delta R  < 0.3$ with respect to the muon candidate, and with a
distance of closest approach with respect to the muon candidate below 1\mm, and divides this sum by the muon
candidate \pt.  The largest value of the isolation parameter among the three muons is used by the BDT\@.

The BDT is
trained and tested on independent samples with no evidence of overtraining.  The BDT output was also verified to be
independent of the trimuon invariant mass.  A BDT for the normalization mode is similarly
trained using the same 10 variables (modified to account for one less muon).  The efficiency as a function of the BDT
requirement is measured with both actual and simulated data for the normalization mode.  The largest discrepancy, 5\%,
is taken as a systematic uncertainty associated with modeling the BDT efficiency.

\begin{figure}[hbtp]
\centering
\includegraphics[width=0.45\textwidth]{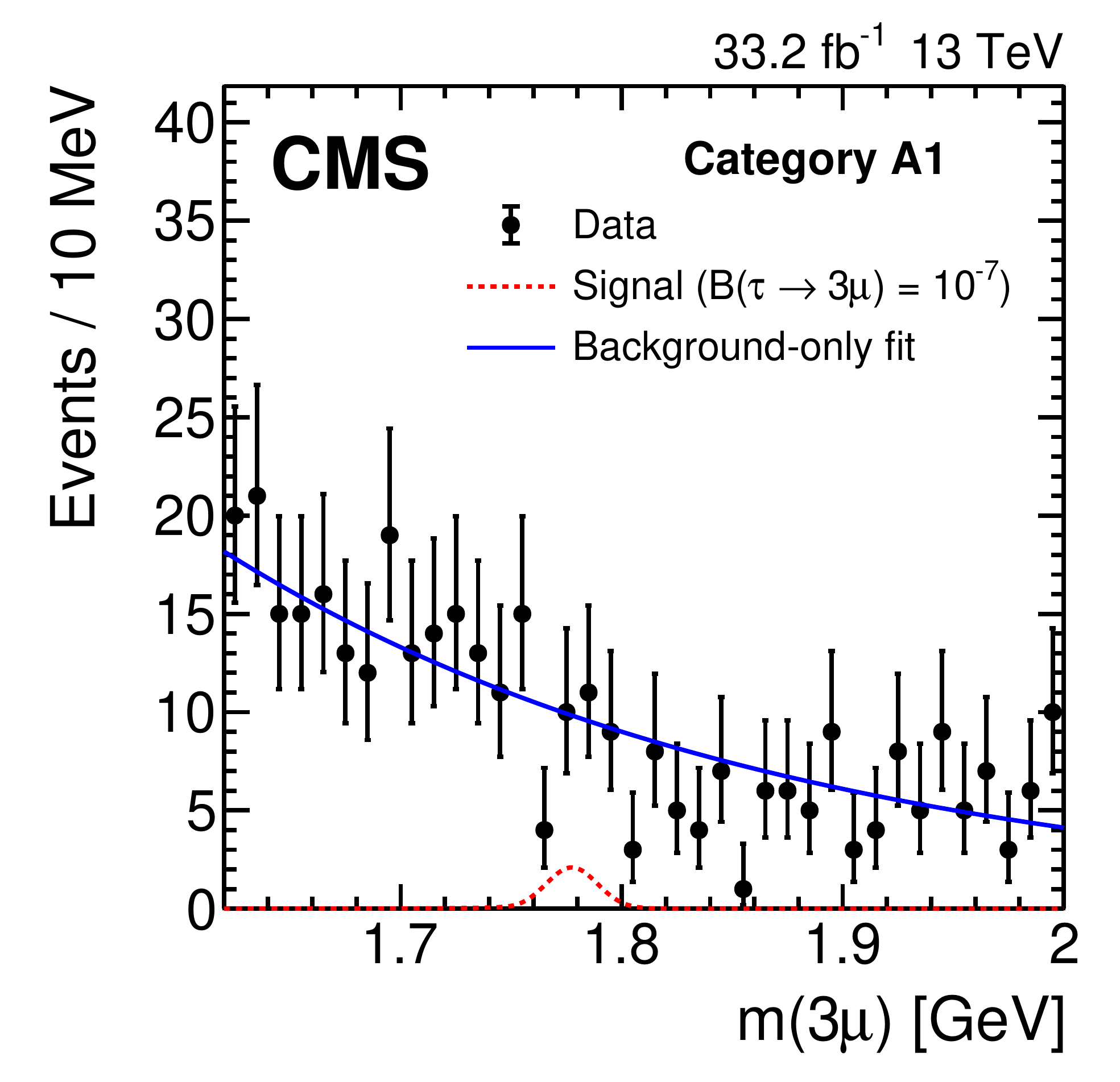}
\includegraphics[width=0.45\textwidth]{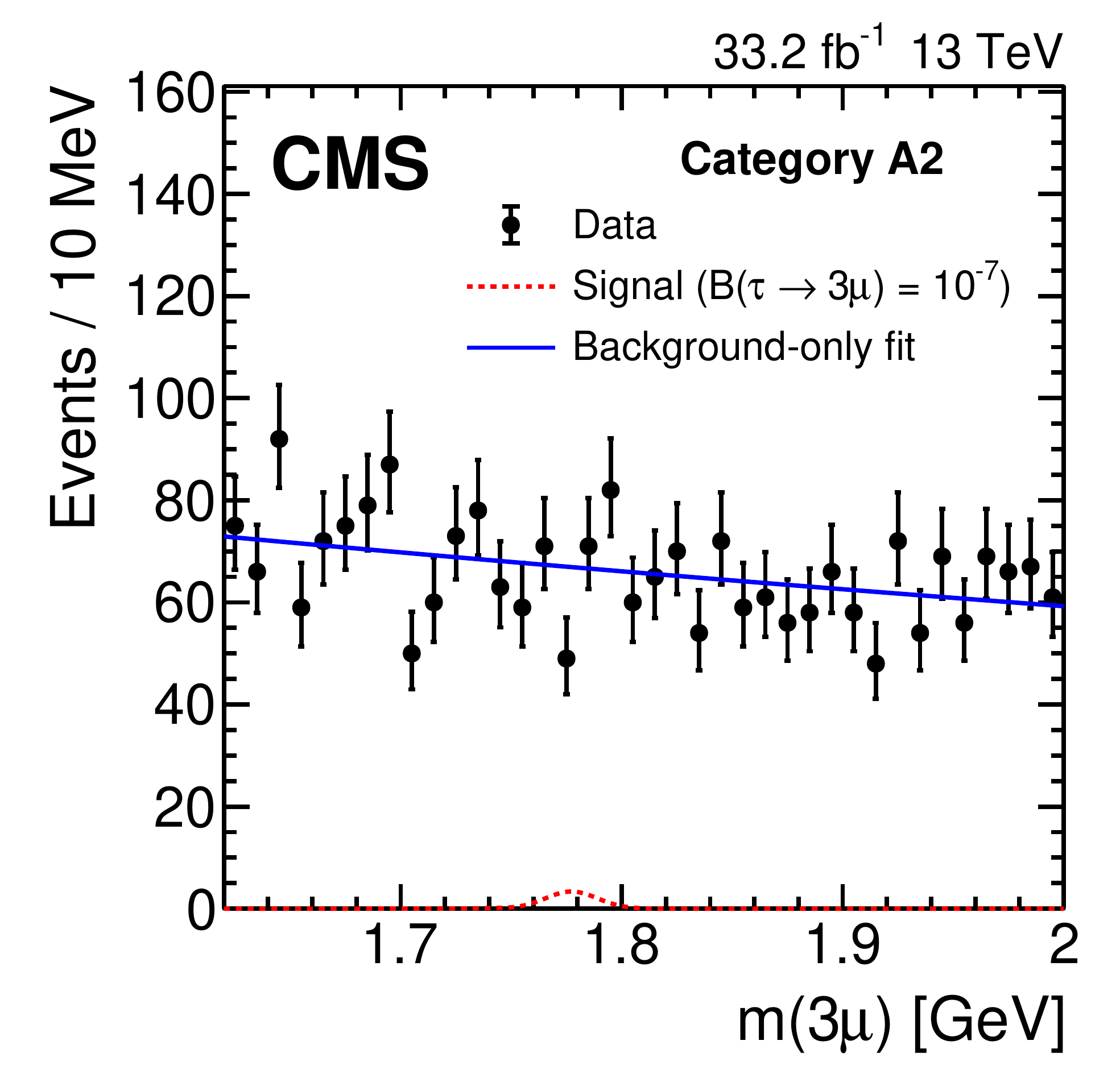}\\
\includegraphics[width=0.45\textwidth]{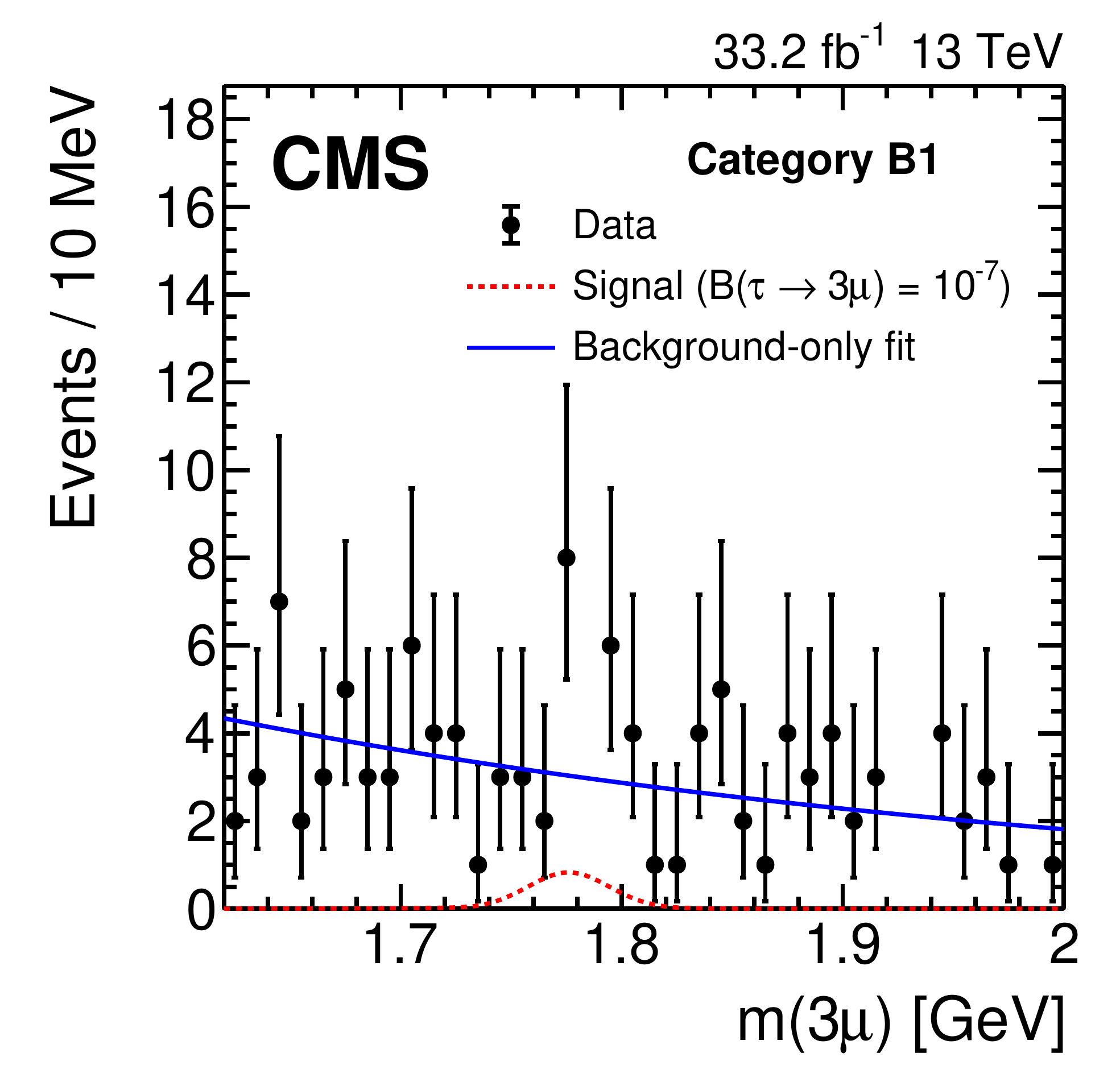}
\includegraphics[width=0.45\textwidth]{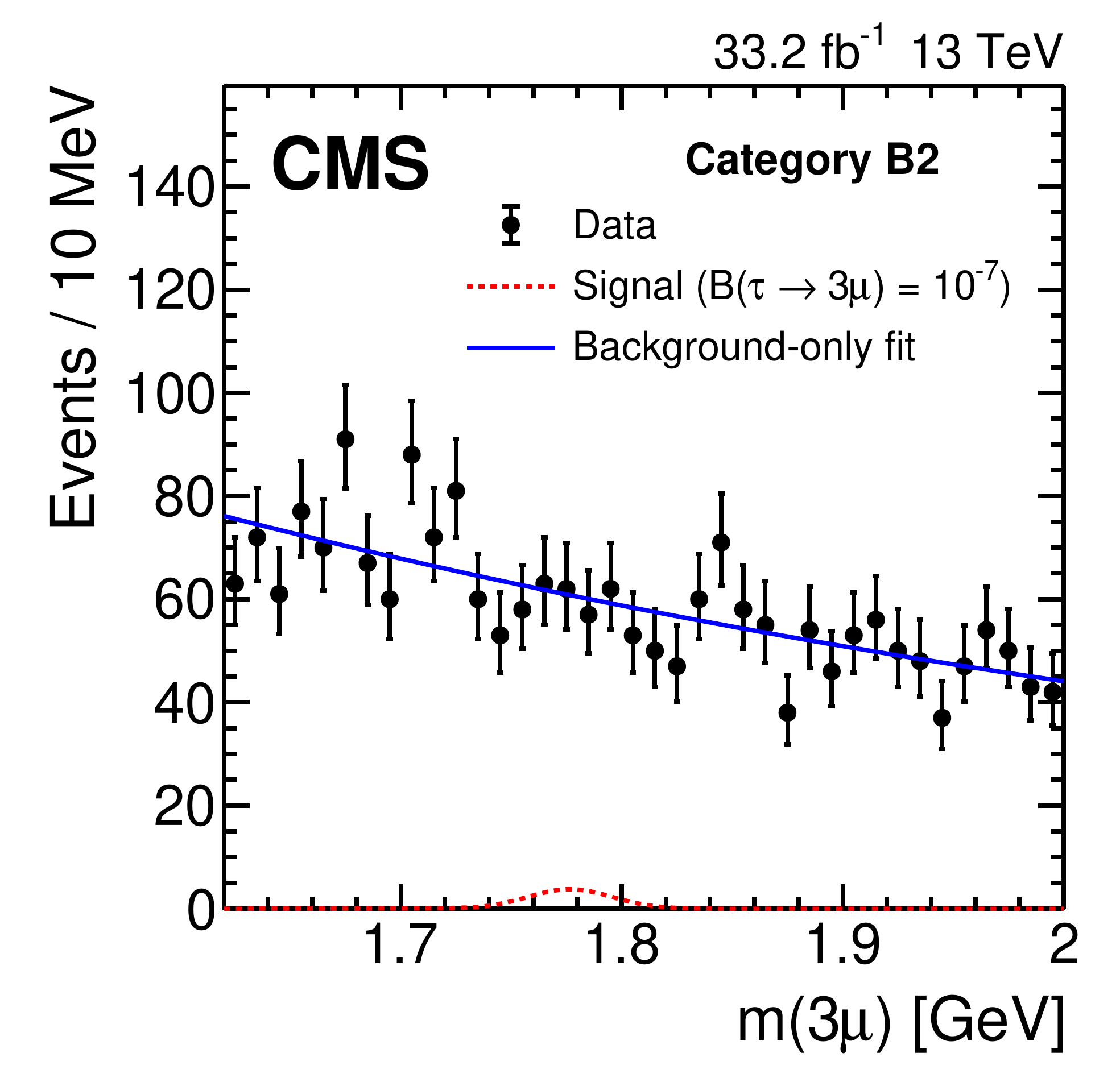}\\
\includegraphics[width=0.45\textwidth]{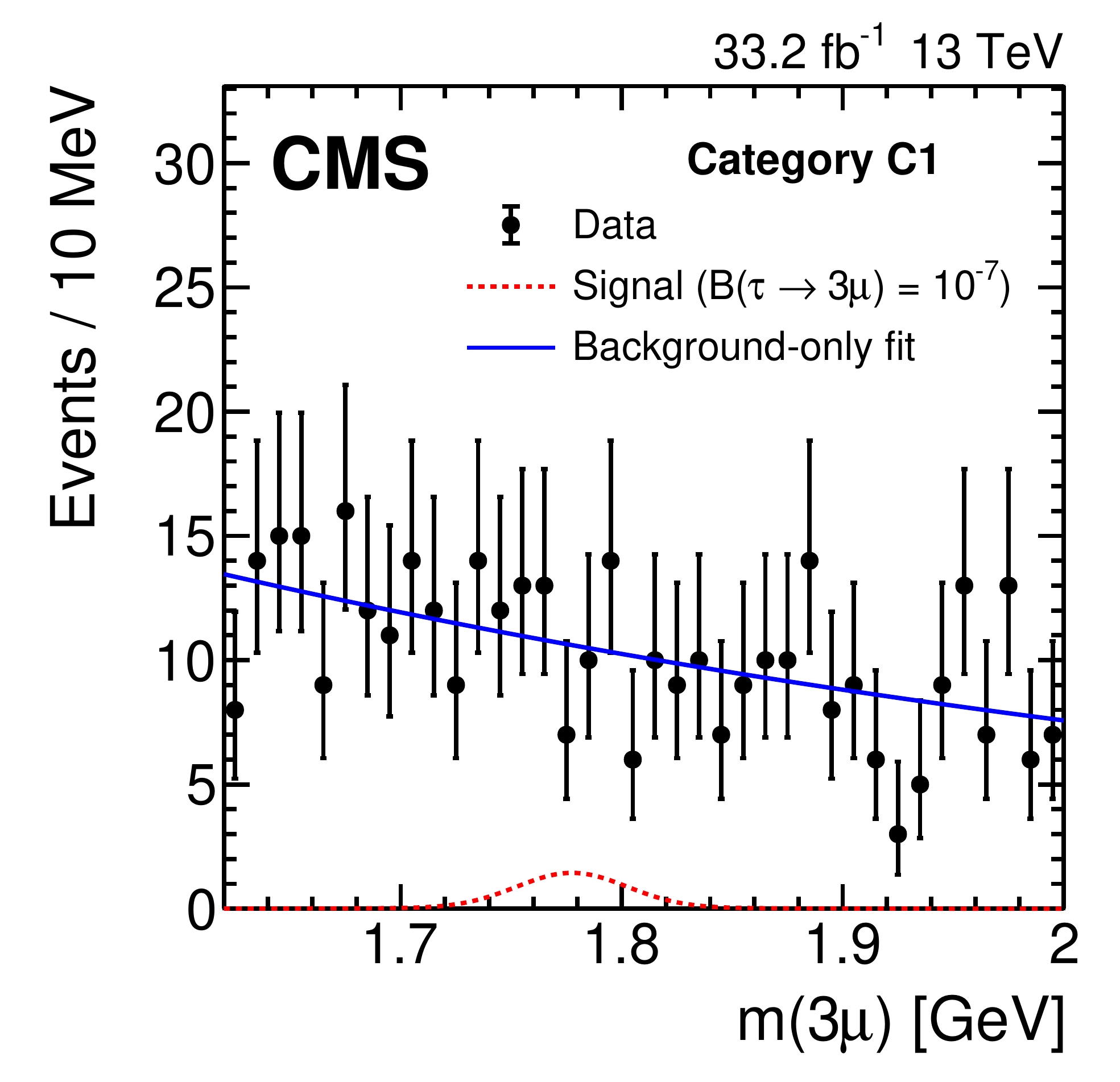}
\includegraphics[width=0.45\textwidth]{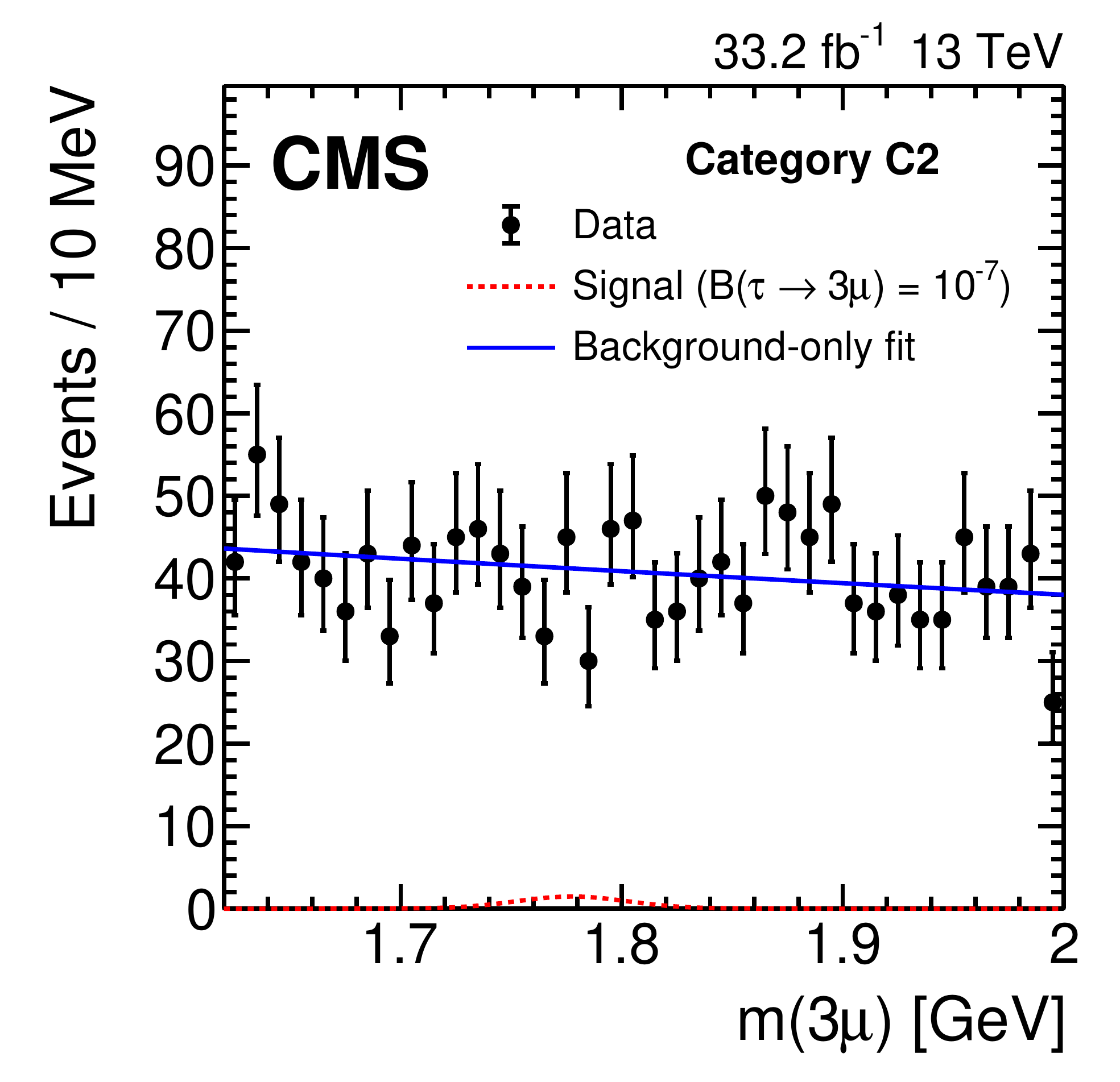}
\caption{Trimuon invariant mass distributions in the six independent event categories used in the heavy-flavor analysis and defined in
the text: A1, A2, B1, B2, C1, C2. The data are shown with filled circles and vertical bars representing the statistical
uncertainty.  The background-only fit and the expected signal for $\mathcal{B}(\tautrimu) = 10^{-7}$ are shown with
solid and dashed lines, respectively.}
\label{fig:HF_massplots}
\end{figure}

To improve the sensitivity of the analysis, the \PGt candidates are separated into six categories depending on the BDT
score and the trimuon invariant mass resolution (the ratio of the mass uncertainty $\sigma_m$, calculated from propagating the
track parameter uncertainties, to the invariant mass $m$).  There are three
mass resolution bins: $\sigma_m/m \leq 0.7\%$, $0.7\% < \sigma_m/m < 1.0\%$, and $\sigma_m/m \geq 1.0\%$, with average mass
resolutions of 12, 19, and 25\MeV, and labeled A, B, and C,
respectively.  The first and last bins roughly correspond to barrel and endcap events, respectively.  Each mass
resolution bin is then divided into three bins based on the BDT score.
The highest two BDT bins in each mass resolution bin are used in the search, with the highest signal-to-background bin
given the label ``1'' and the other ``2''.  Thus, the six categories are labeled A1, A2, B1, B2, C1, and C2.  The
values of the two BDT bin boundaries in each mass resolution bin are determined independently by simultaneously scanning both values
to find the result that gives the best expected upper limit on $\mathcal{B}(\tautrimu)$. 
The trimuon invariant mass distribution for each category is shown in Fig.~\ref{fig:HF_massplots}, along with a
background-only fit (described in Section~\ref{sec:results}) and the contribution expected for a signal with
$\mathcal{B}(\tautrimu)=10^{-7}$.

\subsection{Signal yield normalization}
\label{ssec:HFnormalizing}

Results from simulation indicate that the \PGt leptons in the data sample overwhelmingly come from three disjoint sources: prompt \PD meson decays
(the \PD meson is not from a \PQb hadron decay), \PQb hadron decays (directly from \PQb hadron decays), and nonprompt \PD meson decays (the \PD is from
a \PQb hadron decay), with contributions of 65, 25, and 10\%, respectively.  More than 95\% of the \PGt leptons produced
from charm meson decays are from \PsDp meson decays, with the remainder from \PDp meson decays.  Approximately 75\% of the signal is
expected to come from the L1 dimuon trigger, and can be directly calibrated using
\dsphipi events since they pass the same trigger.  The remaining 25\% of the expected signal is
obtained exclusively from the L1 trimuon trigger.   As
detailed in Section~\ref{sec:results}, the final results are obtained from a fit that uses both the expected number
of background events and the relationship between $\mathcal{B}(\tautrimu)$ and the expected number of signal events.
While this relationship can be obtained from an equation similar to Eq.~(\ref{eq:wbr}), the heavy-flavor production cross sections
have large uncertainties.  To mitigate this, and correct for effects like the L1 trigger misconfiguration during the first half of 2016,
we extract the expected signal yields using methods based on control samples in data to calibrate the production of \PGt leptons.

\subsubsection{Yield of events from dimuon L1 triggers}

\begin{figure}[hbtp]
\centering
\includegraphics[width=0.45\textwidth]{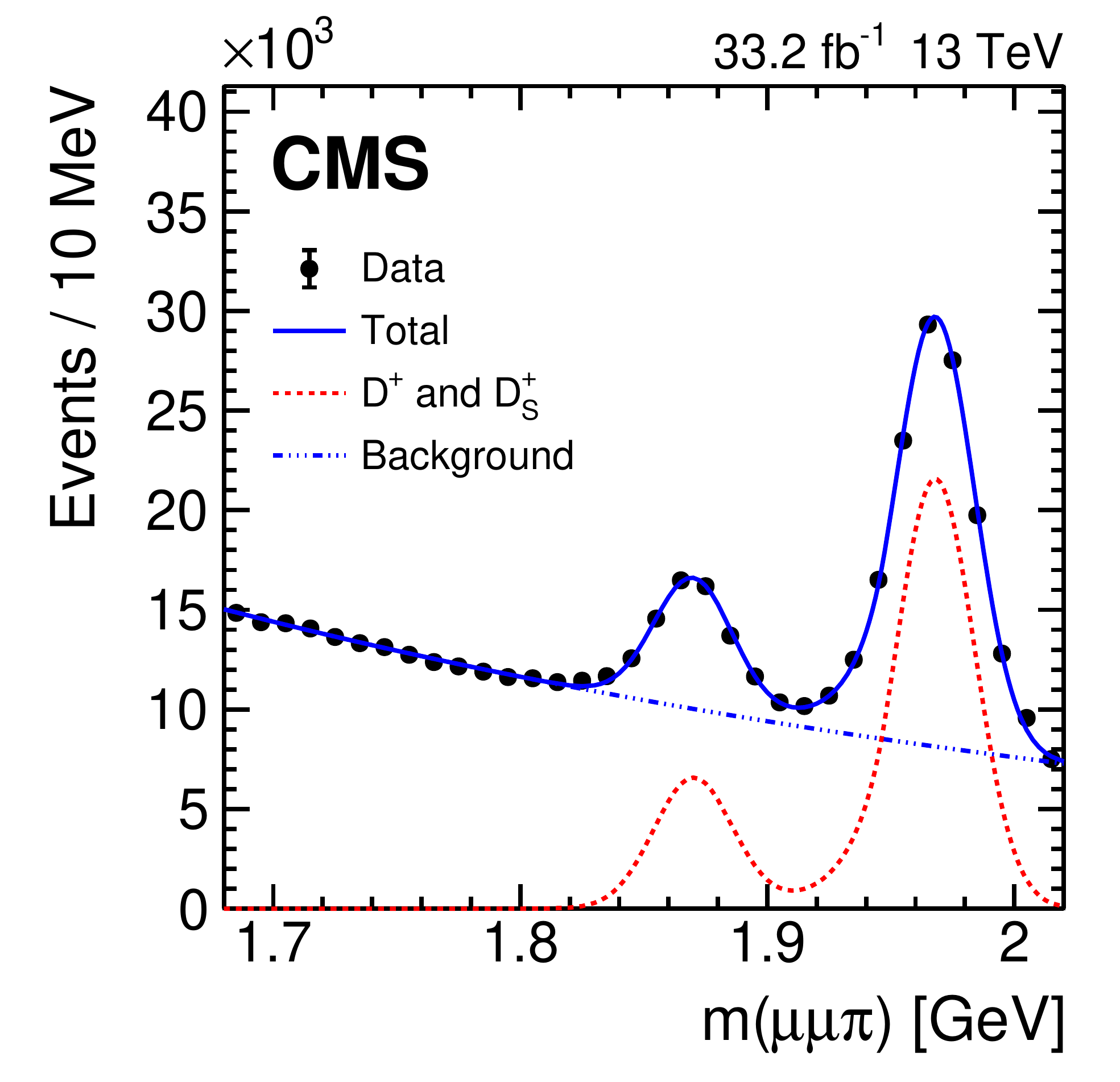}
\includegraphics[width=0.45\textwidth]{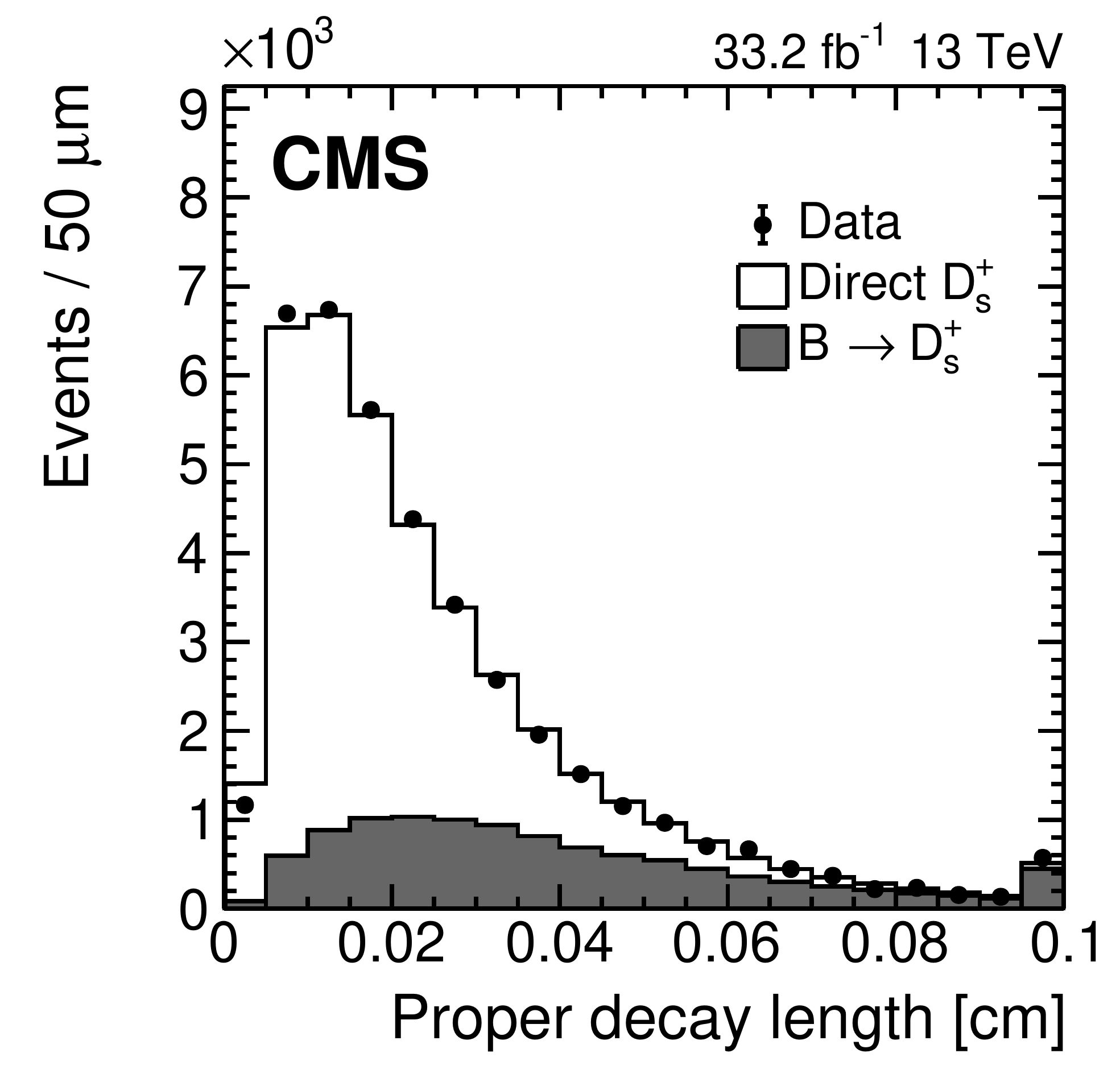}
\caption{Left: the $\PGm\PGm\Pgp$ invariant mass distribution with fits to the \PDp and \PsDp peaks and the background.  Right: background-subtracted
proper decay length distribution for \dsphipi
events (points) and the fitted contributions from \PsDp mesons produced directly (open histogram) and from \PQb hadron
decays (filled histogram).  The highest bin also contains the overflow events.  The vertical bars in both plots represent the statistical uncertainties
in the data.}
\label{fig:DtoPhiPi}
\end{figure}

The expected number of \tautrimu signal events from \PsDp meson decays that pass the dimuon L1 triggers, denoted $N_{\text{sig}(\PD)}$, is related to $\mathcal{B}(\tautrimu)$ by:
\begin{linenomath}
\begin{equation}
\label{eq:Ds_contribution}
N_{\text{sig}(\PD)} = N_\text{norm}
\frac{\mathcal{B}(\dstaunu)}{\mathcal{B}(\dsphipi)} 
\frac{\mathcal{A}_{3\Pgm(\PD)}}{\mathcal{A}_{\Pgm\Pgm\Pgp}}
\frac{\epsilon_{3\PGm(\PD)}^{\text{reco}}}{\epsilon_{\Pgm\Pgm\Pgp}^{\text{reco}}}
\frac{\epsilon_{3\PGm(\PD)}^{2\PGm\text{trig}}}{\epsilon_{\Pgm\Pgm\Pgp}^{2\Pgm\text{trig}}}\,
\mathcal{B}(\tautrimu),
\end{equation}
\end{linenomath}
where $N_\text{norm}$ is the measured \dsphipi yield, $\mathcal{A}$,
$\epsilon^\text{reco}$, and $\epsilon^\text{trig}$ are the detector acceptance, selection efficiency, and trigger
efficiency for the two channels, respectively, and the branching fractions are $\mathcal{B}(\dstaunu)=(5.48\pm0.23)\%$ and
$\mathcal{B}(\dsphipi) = (1.3\pm0.1)\times 10^{-5}$~\cite{PDG2018}.
Figure~\ref{fig:DtoPhiPi} (left) shows the $\PGm\PGm\Pgp$ invariant mass distribution with fits to the \PDp and \PsDp peaks using
Crystal Ball functions~\cite{CrystalBallRef} for the signal and an exponential function for the background, from which $N_\text{norm}$ can
be extracted from the peak on the right.  Note that $N_{\text{sig}(\PD)}$ includes contributions from directly produced \PsDp
mesons and \PsDp mesons from \PQb hadron decays.  To evaluate the degree to which the normalization mode mimics the signal
mode, the ratio of the \dsphipi yield to the number of signal sideband events is
measured for seven different run periods. Assuming these seven values are measuring the same quantity, we use the
scale-factor method~\cite{PDG2018} to derive a systematic uncertainty of 10\%.

The expected number of \tautrimu signal events from decays of the form \BtauX coming from the dimuon L1 triggers, denoted $N_{\text{sig}(\PB)}$,
is related to $\mathcal{B}(\tautrimu)$ by:
\begin{linenomath}
\begin{equation}
\label{eq:B_contribution}
N_{\text{sig}(\PB)} = N_\text{norm}\, f\, 
\frac{\mathcal{B}(\BtauX)}{\mathcal{B}(\dsphipi) \mathcal{B}(\BDsX)} 
\frac{\mathcal{A}_{3\PGm(\PB)}}{\mathcal{A}_{\Pgm\Pgm\Pgp}}
\frac{\epsilon_{3\PGm(\PB)}^{\text{reco}}}{\epsilon_{\Pgm\Pgm\Pgp}^{\text{reco}}}
\frac{\epsilon_{3\PGm(\PB)}^{2\PGm\text{trig}}}{\epsilon_{\Pgm\Pgm\Pgp}^{2\Pgm\text{trig}}}\,
\mathcal{B}(\tautrimu),
\end{equation}
\end{linenomath}
where $N_\text{norm}$ is the measured \dsphipi yield, $f$ is the fraction of observed
$\PsDp$ mesons from \PQb hadron decays, and $\mathcal{A}$, $\epsilon^\text{reco}$, and $\epsilon^\text{trig}$ are the
detector acceptance, selection efficiency, and trigger efficiency for the two channels, respectively.  The newly introduced
branching fractions are $\mathcal{B}(\BtauX)=(3.4\pm0.4)\%$ (including the measured 2.7\% from
$\PB \!\to\! \PGt\PGn\PD^{(*)}$ decays~\cite{PDG2018} and an estimated 0.7\% from other decays based on
\PYTHIA) and $\mathcal{B}(\BDsX) = (10.0\pm1.6)\%$ (averaging the measured \PBz and \PBm branching
fractions~\cite{PDG2018}).

The fraction $f$ can be calculated as
$f = \sigma(\Pp\Pp\!\to\!\PB)\mathcal{B}(\BDsX)/\sigma(\Pp\Pp\!\to\!\PsDp)$.  Since the \PsDp mesons produced from
\PQb hadron decays will tend to decay farther from the $\Pp\Pp$ collision vertex than directly produced \PsDp mesons, we
use the proper decay length distribution to measure $f$.  The proper decay length is $LM/p$ where $L$ is the
distance between the primary vertex and the $\Pgm\Pgm\Pgp$ vertex, $M$ is the $\Pgm\Pgm\Pgp$ invariant mass, and $p$ is
the $\Pgm\Pgm\Pgp$ momentum.  Figure~\ref{fig:DtoPhiPi} (right) shows the proper decay length distribution for \PsDp
mesons in which the background has been subtracted using the invariant mass sidebands.  The proper decay length distribution shapes
for \PsDp mesons directly produced (open histogram) and from \PB decays (shaded histogram) are obtained from simulation.
The data distribution is fit to a linear sum of these two simulation shapes, yielding a measured value of $f=0.267 \pm 0.015$.
The value from simulation of $0.240\pm 0.001$ is used in the analysis and the difference between the two values is included as a systematic uncertainty.

The small contributions from $\PDp\!\to\!\PGt+X$ and $\PBsz \!\to\! \PGt+X$ decays are added by scaling the \dstaunu and
\BtauX predictions by 0.04 and 0.12, respectively, as determined from simulation.  A systematic uncertainty
equal to the total contribution in each case is assessed.  The much smaller contribution $({\sim}0.1\%)$ from direct
\PQb baryon decays is not included.

Uncertainties in the ratios of event selection acceptances $(\mathcal{A}_{3\Pgm(\PD)}/\mathcal{A}_{\Pgm\Pgm\Pgp}$ and
$\mathcal{A}_{3\PGm(\PB)}/\mathcal{A}_{\Pgm\Pgm\Pgp})$ are estimated by changing the parton distribution function sets
in the corresponding simulated events.  Although the acceptances change by up to 7\%, the ratios remain constant
within $\mathcal{O}(1\%)$, consistent with the statistical uncertainty associated with the size of the simulated
samples.  In the ratio $\epsilon_{3\PGm}^{\text{reco}}/\epsilon_{\PGm\PGm\pi}^{\text{reco}}$,
the muon reconstruction efficiency does not cancel exactly since the numerator refers to events with three muons and the denominator
to events with only two.  We derive data-to-simulation corrections for the muon reconstruction efficiency in bins of
muon \pt and $\eta$ using the tag-and-probe method~\cite{CMS:2011aa} applied to $\cPJgy \!\to\! \PGm\PGm$ data events.  These
additional corrections are then applied to signal events. The systematic uncertainty in the correction is estimated to
be 1.5\%.

\subsubsection{Yield of events exclusively from trimuon L1 triggers}

As described in Section~\ref{sec:selection}, the data are collected using both dimuon and trimuon triggers.  The data
collected using trimuon triggers cannot be directly normalized to \dsphipi, as this
decay only contains two muons. 
The simulation predicts that the fraction of signal events triggered exclusively through the L1 trimuon trigger is
33\% of the events passing the L1 dimuon triggers.  When measured from events in the sideband region, this
ratio is found to be 35\% using data collected after the initial trigger problems were fixed, a 6\% difference.  The
data-to-simulation correction for the dimuon trigger, measured in \dsphipi events, is
0.90 for the same data-taking period.  We scale up the dimuon-triggered predicted yields for this data-taking
period by the simulation value of 33\% and assign a systematic uncertainty of 12\% to account for the observed
6 and 10\% differences.  For the initial data-taking periods, the expected yield is scaled by the ratio of the trimuon
trigger rates in the early and late periods, with the same 12\% uncertainty.

\subsection{Systematic uncertainties}
\label{ssec:HFsystematics}

The systematic uncertainties associated with the expected signal event yield,
as described previously, are summarized in Table~\ref{tab:HFsystematics}.
In addition, systematic uncertainties related to the signal and background shapes are evaluated.
The signal invariant mass shape uncertainties are estimated by comparing data and simulation results for the fitted
value of the mean and resolution in \dsphipi decays.  The mean value is found to be
0.07\% higher in simulation and therefore the mass in the signal simulation is shifted by ${-}0.07\%$ with a systematic uncertainty of
0.07\%.  The resolution is found to be 2\% smaller in simulation and thus the signal simulation resolution is increased by 2\%,
with a systematic uncertainty of 2.5\%, consistent with the statistical precision of the measurement.  The uncertainty
in the background shape is obtained by varying the functional form from the default exponential to a third-order
polynomial and a power-law function.  This is found to contribute an uncertainty of less than 1\%.

\begin{table}[htbp]
\centering
\topcaption{The sources of systematic uncertainties in the heavy-flavor analysis affecting signal modeling and their impact on 
the expected signal event yield.  The columns labeled Uncertainty and Yield give the relative uncertainty associated with the source, and the
resulting effect on the yield, respectively.}
\begin{tabular}{lccc}
Source of uncertainty                      & Uncertainty (\%) & Yield (\%) \\
\hline
\PsDp normalization  & 10         & 10   \\
$\mathcal{B}(\dstaunu)$ & 4                  & 3    \\
$\mathcal{B}(\dsphipi)$ & 8     & 8   \\
$\mathcal{B}(\BDsX)$  & 16                & 5    \\
$\mathcal{B}(\BtauX)$ & 11                & 3    \\
$\PB/\PD$ ratio $f$  & 11                             & 3   \\
Number of events from L1 trimuon trigger  & 12          & 3   \\
Acceptance ratio $\mathcal{A}_{3\PGm}/\mathcal{A}_{\Pgm\Pgm\Pgp}$ & 1          & 1  \\
Muon reconstruction efficiency & 1                                        & 1  \\
BDT requirement efficiency & 5                                                       & 5  \\
\hline
Total                      &                                                        & 16  \\
\end{tabular}
\label{tab:HFsystematics}
\end{table}

\section{Results}
\label{sec:results}

The branching fraction $\mathcal{B}(\tautrimu)$ is extracted from a simultaneous unbinned maximum
likelihood fit to the trimuon invariant mass distribution (1.6--2\GeV) in the two categories of the \PW boson analysis and the six categories of the heavy-flavor
analysis.

For the \PW boson analysis, the signal model is a Gaussian function with fixed mean and width, as determined from fitting the
simulated events in the appropriate category. For the heavy-flavor selection, the signal model is a Gaussian plus
Crystal Ball function~\cite{CrystalBallRef} with fixed mean and width, as determined from fitting the simulated events
in the appropriate category and modified as described in Section~\ref{ssec:HFsystematics}.  In all cases, the background
model is an exponential function with parameters and normalization determined by the fit.

As can be seen in the trimuon invariant mass plots of Figs.~\ref{fig:W_massplots} and \ref{fig:HF_massplots}, no evidence for a \tautrimu signal is found.
Upper limits on $\mathcal{B}(\tautrimu)$ are determined from a fully frequentist
method~\cite{CMS-NOTE-2011-005} based on modified profile likelihood test statistics and the \CLs criterion~\cite{Junk:1999kv,Read:2002hq}. Systematic uncertainties are incorporated in the analysis via nuisance parameters.
Uncertainties are assumed to be uncorrelated
between the two channels. A log-normal probability density function is assumed for the nuisance parameters affecting the
corrected signal yields.  Events from data and simulation that pass the selection criteria of both analyses are removed
from the heavy-flavor analysis in the combined fit.

The observed (expected) upper limit at 90\% \CL on $\mathcal{B}(\tautrimu)$ using all events is $8.0 \times 10^{-8}$
$(6.9 \times 10^{-8})$. Fitting the \PW boson and heavy-flavor events separately returns observed (expected) 90\% \CL
upper limits of $20 \times 10^{-8}$ $(13 \times 10^{-8})$ and $9.2 \times 10^{-8}$ $(10.0 \times 10^{-8})$,
respectively.

\section{Summary}
\label{sec:summary}
The results of a search for the lepton flavor violating decay $\tautrimu$, using proton-proton collisions with a
center-of-mass energy of 13\TeV at the LHC, are presented.  The search uses data collected by CMS in 2016, corresponding to
an integrated luminosity of 33.2\fbinv, and, for the first time, combines the result of two analyses: one targeting \PGt leptons produced
in \PW boson decays and the other using \PGt leptons from heavy-flavor hadron decays.  No signal is observed, and the
branching fraction $\mathcal{B}(\tautrimu)$ is determined to be less than $8.0\times 10^{-8}$ at 90\% confidence
level, with an expected upper limit of $6.9\times 10^{-8}$.  While the limit obtained in this measurement is still a factor
of four away from the current most restrictive one from the Belle experiment~\cite{Hayasaka:2010np}, we have achieved
similar sensitivity to that by BaBar~\cite{Aubert:2007pw} and LHCb~\cite{Aaij:2014azz}.

\begin{acknowledgments}
  We congratulate our colleagues in the CERN accelerator departments for the excellent performance of the LHC and thank the technical and administrative staffs at CERN and at other CMS institutes for their contributions to the success of the CMS effort. In addition, we gratefully acknowledge the computing centers and personnel of the Worldwide LHC Computing Grid for delivering so effectively the computing infrastructure essential to our analyses. Finally, we acknowledge the enduring support for the construction and operation of the LHC and the CMS detector provided by the following funding agencies: BMBWF and FWF (Austria); FNRS and FWO (Belgium); CNPq, CAPES, FAPERJ, FAPERGS, and FAPESP (Brazil); MES (Bulgaria); CERN; CAS, MoST, and NSFC (China); COLCIENCIAS (Colombia); MSES and CSF (Croatia); RIF (Cyprus); SENESCYT (Ecuador); MoER, ERC IUT, PUT and ERDF (Estonia); Academy of Finland, MEC, and HIP (Finland); CEA and CNRS/IN2P3 (France); BMBF, DFG, and HGF (Germany); GSRT (Greece); NKFIA (Hungary); DAE and DST (India); IPM (Iran); SFI (Ireland); INFN (Italy); MSIP and NRF (Republic of Korea); MES (Latvia); LAS (Lithuania); MOE and UM (Malaysia); BUAP, CINVESTAV, CONACYT, LNS, SEP, and UASLP-FAI (Mexico); MOS (Montenegro); MBIE (New Zealand); PAEC (Pakistan); MSHE and NSC (Poland); FCT (Portugal); JINR (Dubna); MON, RosAtom, RAS, RFBR, and NRC KI (Russia); MESTD (Serbia); SEIDI, CPAN, PCTI, and FEDER (Spain); MOSTR (Sri Lanka); Swiss Funding Agencies (Switzerland); MST (Taipei); ThEPCenter, IPST, STAR, and NSTDA (Thailand); TUBITAK and TAEK (Turkey); NASU (Ukraine); STFC (United Kingdom); DOE and NSF (USA).
   
  \hyphenation{Rachada-pisek} Individuals have received support from the Marie-Curie program and the European Research Council and Horizon 2020 Grant, contract Nos.\ 675440, 752730, and 765710 (European Union); the Leventis Foundation; the A.P.\ Sloan Foundation; the Alexander von Humboldt Foundation; the Belgian Federal Science Policy Office; the Fonds pour la Formation \`a la Recherche dans l'Industrie et dans l'Agriculture (FRIA-Belgium); the Agentschap voor Innovatie door Wetenschap en Technologie (IWT-Belgium); the F.R.S.-FNRS and FWO (Belgium) under the ``Excellence of Science -- EOS" -- be.h project n.\ 30820817; the Beijing Municipal Science \& Technology Commission, No. Z191100007219010; the Ministry of Education, Youth and Sports (MEYS) of the Czech Republic; the Deutsche Forschungsgemeinschaft (DFG) under Germany's Excellence Strategy -- EXC 2121 ``Quantum Universe" -- 390833306; the Lend\"ulet (``Momentum") Program and the J\'anos Bolyai Research Scholarship of the Hungarian Academy of Sciences, the New National Excellence Program \'UNKP, the NKFIA research grants 123842, 123959, 124845, 124850, 125105, 128713, 128786, and 129058 (Hungary); the Council of Science and Industrial Research, India; the HOMING PLUS program of the Foundation for Polish Science, cofinanced from European Union, Regional Development Fund, the Mobility Plus program of the Ministry of Science and Higher Education, the National Science Center (Poland), contracts Harmonia 2014/14/M/ST2/00428, Opus 2014/13/B/ST2/02543, 2014/15/B/ST2/03998, and 2015/19/B/ST2/02861, Sonata-bis 2012/07/E/ST2/01406; the National Priorities Research Program by Qatar National Research Fund; the Ministry of Science and Higher Education, project no. 02.a03.21.0005 (Russia); the Programa Estatal de Fomento de la Investigaci{\'o}n Cient{\'i}fica y T{\'e}cnica de Excelencia Mar\'{\i}a de Maeztu, grant MDM-2015-0509 and the Programa Severo Ochoa del Principado de Asturias; the Thalis and Aristeia programs cofinanced by EU-ESF and the Greek NSRF; the Rachadapisek Sompot Fund for Postdoctoral Fellowship, Chulalongkorn University and the Chulalongkorn Academic into Its 2nd Century Project Advancement Project (Thailand); the Kavli Foundation; the Nvidia Corporation; the SuperMicro Corporation; the Welch Foundation, contract C-1845; and the Weston Havens Foundation (USA).
\end{acknowledgments}

\bibliography{auto_generated}

\cleardoublepage \appendix\section{The CMS Collaboration \label{app:collab}}\begin{sloppypar}\hyphenpenalty=5000\widowpenalty=500\clubpenalty=5000\vskip\cmsinstskip
\textbf{Yerevan Physics Institute, Yerevan, Armenia}\\*[0pt]
A.M.~Sirunyan$^{\textrm{\dag}}$, A.~Tumasyan
\vskip\cmsinstskip
\textbf{Institut f\"{u}r Hochenergiephysik, Wien, Austria}\\*[0pt]
W.~Adam, F.~Ambrogi, T.~Bergauer, M.~Dragicevic, J.~Er\"{o}, A.~Escalante~Del~Valle, R.~Fr\"{u}hwirth\cmsAuthorMark{1}, M.~Jeitler\cmsAuthorMark{1}, N.~Krammer, L.~Lechner, D.~Liko, T.~Madlener, I.~Mikulec, F.M.~Pitters, N.~Rad, J.~Schieck\cmsAuthorMark{1}, R.~Sch\"{o}fbeck, M.~Spanring, S.~Templ, W.~Waltenberger, C.-E.~Wulz\cmsAuthorMark{1}, M.~Zarucki
% ...
\vskip\cmsinstskip
\dag: Deceased\\
1:  Also at Vienna University of Technology, Vienna, Austria\\
% ...
\end{sloppypar}
\end{document}